\documentclass[12pt]{article}
\newlength{\vshift}
\input epsf.tex
\newlength{\hshift}

\usepackage{amssymb}
\usepackage{amsmath}
\usepackage{amsthm}
\usepackage{amsopn}
\usepackage{ulem}
\usepackage{cancel}
\usepackage{color}

\usepackage{axodraw4j}
\usepackage{pstricks}
\usepackage{color}

\def\comment#1{}
\usepackage{graphicx}

\begin{document}

 \vspace*{3cm}

\begin{center}
{\large \bf{Neutrino oscillations in nuclear media}}

\vskip 4em

{
 {\bf Iman Motie$^{a,b}$}\footnote{e-mail: i.motie@ph.iut.ac.ir} and {\bf She-Sheng Xue$^{a}$}\footnote{e-mail: xue@icra.it } }
 \vskip 1em {\it a) ICRANet, P.zza della Repubblica 10, I-65122
Pescara, \& Physics Department, University of Rome ``La
Sapienza'', Rome Italy\\} {\it b) Department of Physics, Isfahan
University of Technology, Isfahan 84156-83111, Iran }
\end{center}

 \vspace*{1.5cm}

\begin{abstract}
On basis of effective interactions of charged lepton and
hadron currents, we obtain an effective interacting Hamiltonian of
neutrinos in nuclear media up to the leading order. Using this
effective Hamiltonian, we study neutrino mixing and oscillations in
nuclear media and strong magnetic fields. We compute neutrino
mixing angle and mass squared difference, and find the pattern of vacuum neutrino
oscillations is modified in magnetized nuclear media. Comparing with the vacuum neutrino oscillation, we find that for high-energy neutrinos, neutrino oscillations are suppressed in the presence of nuclear media. In the general case of neutral nuclear media with the
presence of electrons, we calculate the mixing angle and mass squared
difference, and discuss the resonance and level-crossing in
neutrino oscillations. 

\end{abstract}
\newpage

\section{\large Introduction}
In the standard model (SM) for fundamental particle physics,
neutrinos are massless and left-handed. They are produced via
neutral and charge current interactions. In charge current interaction, each
flavor of physical neutrino is produced together with a corresponding
charged lepton. It was proposed by Pontecorvo \cite{pontecrov}
that each physical neutrino state could be a superposition of
different mass eigenstates, i.e., neutrino mixing and
oscillations. This implies that neutrinos are massive. The recent experiments provide strong evidences
that confirm neutrino flavor oscillations
\cite{exp-solar}-\cite{exp-accel}.

Recently, there are many experimental and theoretical studies
dedicating to neutrino physics, in particular neutrino
oscillations. These studies to know neutrino masses and mixing
angles can be one of the best way for understanding new physics
beyond the Standard Model. This new physics can be related to
fundamental symmetries of theories, 
fundamental dynamics of quantum gravity
and other dynamics in extensions of the standard model. In
addition, the knowledges about neutrino masses, mixing angles and
oscillations are important for studying astrophysics and
cosmology.

On the other hand, neutrino oscillations can significantly be
modified, when neutrinos
travel through media rather than the vacuum. This effect occurs
when neutrinos under consideration experience different
interactions by passing through media. Neutrino oscillations in
media can be large, even although neutrino oscillations in the
vacuum is small. This medium effect was studied by Wolfenstein
\cite{wolf}, Mikheyev and Smirnov  \cite{smirniv} (MSW), which we
briefly discuss in Sec.~(\ref{uniform1}).

In this article, by considering the neutrino-lepton current
interacting hadronic current, we study the nuclear effect on
neutrino oscillations while neutrinos are propagating in nuclear
media and strong magnetic fields. In the system of two flavor
neutrinos, we compute the neutrino mass squared difference, mixing
angle and neutrino oscillation probability, and we show the
pattern of neutrino vacuum oscillations is modified by effects of
nuclear media and strong magnetic fields. The resonance of
neutrino oscillations probability and the inversion of neutrino
flavors are discussed. We show that for high-energy neutrinos, the effect of nuclear media on neutrino oscillations is important, neutrino flavor oscillations are suppressed. This should be considered in studying properties of neutrinos produced inside neutron stars, quark stars and magnetars in astrophysics.   

This article is arranged as follows. In Secs.~\ref{vaco} and
\ref{uniform1}, we briefly discuss neutrino oscillations in the
vacuum and normal media. In Sec.~\ref{nnint} we
describe neutrino scattering in nuclear media by effective
current-current interactions in the SM. In Sec.~\ref{nuco}, we present our
results of neutrino oscillations in nuclear media. The case of strongly
magnetized nuclear media is considered in Sec.~\ref{mag}. In
Sec.~\ref{g} we consider neutrino oscillation in the general case of
neutral nuclear media with the presence of electron and strong magnetic fields.
The neutrino oscillation resonance and flavor inversion are
studied in Sec.~\ref{reson}. The last section, summary and
remarks are given.

\section{\large Neutrino oscillation in vacuum} \label{vaco}

Flavor neutrinos $(\nu_e,\nu_\mu, \nu_\tau)$ are always produced
and detected in flavor eigenstates via their interacting with
intermediate gauge bosons $W^{(\pm)}_\mu$ and $Z_\mu^0$ in the SM.
Flavor neutrinos are weak interaction eigenstates and in principle they are
superpositions of the mass
eigenstates $|\nu_i\rangle$
\begin{eqnarray}
\mathcal{H}|\nu_i\rangle&=&E_i|\nu_i\rangle,\quad i=1,2,3,
\label{h0}
\end{eqnarray}
where $E_i$ are the energy eigenvalues of the type-$i$ neutrino. For ultra-relativistic
neutrinos,  neutrino energies can be approximately written as
\begin{eqnarray}
E_i&\approx&p_i+\frac{m^2_i}{2p_i}\,, \label{ei}
\end{eqnarray}
where $m_i$ and $p_i$ are the type-$i$ neutrino
mass and momentum respectively, and $p_i\gg m_i$.

Flavor eigenstates and Hamiltonian
eigenstates (mass eigenstates) are related by an unitary
transformation represented by a matrix $U$,
\begin{eqnarray}
|\nu_l\rangle&=&\sum\limits_{i=1}^{3}U_{li}|\nu_i\rangle
\label{ij},
\end{eqnarray}
where the flavor index $l=e,\mu,\tau$. This shows that flavor
eigenstates is mixing of Hamiltonian eigenstate $|\nu_i\rangle$
and {\it vice versa}. Time evolution of flavor neutrino states are
given by
\begin{eqnarray}
|\nu_l(t)\rangle&=&e^{-i\mathcal{H}t}|\nu_l\rangle=\sum\limits^{3}_{i=1}e^{-iE_it}U_{li}|\nu_i\rangle\,,
\label{newbase0}
\end{eqnarray}
indicating, after some time $t$, the evolution of these flavor states leads to flavor neutrino oscillations.
The probability of such neutrino oscillations is given by
\begin{eqnarray}
P_{\nu_l\rightarrow\nu_{l'}}=|\langle
\nu_{l'}|\nu_{l}\rangle|^2=\sum\limits_{i,j}|U_{li}U^*_{l'i}U^*_{lj}U_{l'j}|
\cos(\frac{\Delta m^2_{ij}}{2E}t+\varphi_{ll'}) \label{prob},
\end{eqnarray}
where $\Delta m^2_{ij}=m^2_i-m^2_j$ and $\varphi_{ll'}={\rm
arg}(U_{li}U^*_{l'i}U^*_{lj}U_{l'j}) \cite{mohapatra}$.

Assume that there are only two flavor neutrino species, for example $\nu_e$ and
$\nu_{\mu}$. The unitary matrix $U$ is explicitly given by
\begin{eqnarray}
U=\left(%
\begin{array}{cc}
  \cos\theta & \sin\theta \\
  -\sin\theta & \cos\theta \\
\end{array}%
\right), \label{uni1}
\end{eqnarray}
where $\theta\equiv\theta_{12}$ presents a mixing angle. Eq.~(\ref{ij}) becomes
\begin{eqnarray}
|\nu_e\rangle&=&\cos\theta|\nu_1\rangle+\sin\theta|\nu_2\rangle\,,\nonumber\\
|\nu_\mu\rangle&=&-\sin\theta|\nu_1\rangle+\cos\theta|\nu_2\rangle\,.
\label{newbase}
\end{eqnarray}
The Hamiltonian (\ref{h0}) in the base of mass eigenstates is
\begin{eqnarray}
\mathcal H_v&=&\left(%
\begin{array}{cc}
  E_1 & 0 \\
  0 & E_2 \\
\end{array}%
\right)\simeq E+\left(%
\begin{array}{cc}
  m^2_1/2E & 0 \\
  0 & m^2_2/2E \\
\end{array}%
\right) \label{hmass},
\end{eqnarray}
where because of $p_i\gg m_i$, based on Eq.~(\ref{ei}) the leading contribution to the neutrino energy $E_i$ is obtained by assuming $p_1\approx p_2=E$. By using
Eqs.~(\ref{uni1},\ref{hmass}) the Hamiltonian in the base of flavor eigenstates is given by
\begin{eqnarray}
\hat{\mathcal H}_v &=& U{\mathcal H}U^\dag\nonumber\\
&=&E+\frac{m_1^2+m_2^2}{4E}+\frac{\Delta m^2}{4E}
\left(%
\begin{array}{cc}
  -\cos2\theta & \sin2\theta \\
  \sin2\theta & \cos2\theta \\
\end{array}%
\right), \label{h-vac}
\end{eqnarray}
where $\Delta m^2\equiv\Delta m^2_{12}=m_2^2-m_1^2\, (m_2>m_1)$ and the mixing angle $\theta$ is given by
\begin{eqnarray}
\tan 2\theta&=&\frac{2\hat{\mathcal H}_{12}}{\hat{\mathcal
H}_{22}-\hat{\mathcal H}_{11}},
\label{theta}
\end{eqnarray}
and the probability of neutrino oscillation is
\begin{eqnarray}
P_{\nu_{\mu}\leftrightarrow\nu_e}(t)&=&\sin^22\theta\sin^2\left[(E_2-E_1)t
\right]
\nonumber\\
&=&\sin^22\theta\sin^2\left[(E_2-E_1)L \right],\label{pro2}
\end{eqnarray}
where $E_2-E_1=\Delta m^2/(2E)$ and the second line is for relativistic neutrinos $L\approx ct$. This is the description of neutrino
flavor mixing and oscillation in the vacuum. The discussions and
calculations are applied for two-level systems
($\theta_{23},\Delta m^2_{23}$) and ($\theta_{13},\Delta
m^2_{13}$). About this section, readers are referred to Refs.~\cite{mohapatra}-\cite{zober} for more details.


\section{\large Neutrino oscillations in normal media}\label{uniform1}

In the presence of a normal medium, effective neutrino mixing angles and
mass squared  differences can be modified by their interacting with particles in medium. Although interacting cross-sections are very small, modifications can be important if particle densities are very large.

Suppose that neutrinos travel through a normal neutral medium, where charged leptons are mainly electrons, the
description of neutrino oscillations (\ref{uni1}-\ref{pro2}) is
modified due to the fact that electron neutrino scattering with electrons in medium are affected by both charged and neutral current interactions, while muon and tau neutrinos scatterings with medium are affected by the neutral current interaction only.
 This difference modifies the neutrino flavor mixing and
oscillation in the vacuum as described in
Eqs.~(\ref{uni1}-\ref{pro2}). In
the base of flavor eigenstates for two neutrino flavors
$\nu_e$ and $\nu_{\mu}$,  the charged and neutral current
interactions in normal media induce an effective potential \cite{wolf,smirniv}
\begin{equation}
V_C = \sqrt{2}\,G_Fn_e\left(%
\begin{array}{cc}
  1 & 0 \\
  0 & 0 \\
\end{array}%
\right),\quad V_N = -\frac{1}{\sqrt{2}}\,G_Fn_n\left(%
\begin{array}{cc}
  1 & 0 \\
  0 & 1 \\
\end{array}%
\right), \label{vc}
\end{equation}
where $n_e$ and $n_n$ are the number densities of electrons and
nucleons. The potential $V_N$ from the neutral-current interaction
is the same for all of neutrino flavors, and it only shifts the
neutrino energy by a negligible small amount and does not affect neutrino oscillations. Due to the additional effective
potentials $V_C$ and $V_N$ (\ref{vc}), the Hamiltonian
(\ref{h-vac}) is changed to
\begin{eqnarray}
\hat{\mathcal H}_m&=& E+\frac{m_1^2+m_2^2}{4E}-\frac{1}{\sqrt{2}}G_Fn_n \nonumber\\&+&\frac{\Delta m^2}{4E}\left(%
\begin{array}{cc}
-\cos2\theta & \sin2\theta\\\!\!
\sin2\theta  & \cos2\theta\end{array}%
\right)+\left(%
\begin{array}{cc}
  \sqrt{2}G_Fn_e & 0 \\
  0 & 0 \\
\end{array}%
\right). \label{h-norm}
\end{eqnarray}
Using the unitary matrix (\ref{uni1}) characterized by $\theta_m$
the mixing angle in normal media to diagonalize this Hamiltonian,
one obtains the effective mass squared difference $\Delta m^2_m$
and the mixing angle $\theta_m$ \cite{wolf,smirniv} (see also
\cite{mohapatra})
\begin{eqnarray}
\Delta m^2_m &=& \sqrt{(\Delta m^2\cos2\theta-2\sqrt{2}G_Fn_eE)^2+(\Delta m^2\sin
2\theta)^2},
\label{media0}\\
\tan 2\theta_m &=& \frac{\Delta m^2\sin 2\theta}{\Delta m^2\cos
2\theta-2\sqrt{2}G_Fn_eE}.
\label{media1}
\end{eqnarray}
In this base of Hamiltonian eigenstates, flavor neutrino states are expressed as
\begin{eqnarray}
|\nu_e\rangle&=&\cos\theta_m|\nu_1\rangle_m+\sin\theta_m|\nu_2\rangle_m\,,
\nonumber\\|\nu_\mu\rangle&=&-\sin\theta_m|\nu_1\rangle_m+\cos\theta_m|\nu_2\rangle_m\, ,
\label{newbasem}
\end{eqnarray}
and the probability of neutrino oscillations in normal media is given,
\begin{eqnarray}
P^m_{\nu_{\mu}\leftrightarrow\nu_e}(t)&=&\sin^22\theta_m\sin^2\left[(E^m_2-E^m_1)t
\right],
\label{pro3}
\end{eqnarray}
where $E^m_2$ and $E^m_1$ are energy eigenvalues of the Hamiltonian (\ref{h-norm}) and
\begin{eqnarray}
E^m_2-E^m_1=\frac{\Delta m^2_m}{2E}. \label{em12}
\end{eqnarray}
In the absence of a normal medium, $V_C=0$, $\Delta m^2_m = \Delta
m^2$ and $\theta_m = \theta$, one obtains neutrino flavors mixing
and oscillations (\ref{pro2}) in the vacuum.
\comment{
Moreover, we discuss
following three cases:
\begin{itemize}
\item small mixing angle citation ???, $m_2\gg m_1$, i.e.,
$m_2^2\sim \Delta m^2$, and $\Delta m^2\ll A$, leading to
$\theta_m \sim 0$ and $\Delta m^2_m\simeq A$, and small neutrino
oscillation probability (\ref{pro3}); \item small mixing angle
citation ???, $m_2\gg m_1$, i.e., $m_2^2\sim \Delta m^2$ and  for
$\Delta m^2\gg A$, leading to $\theta_m \sim \pi/4$, $\Delta
m^2_m\simeq \Delta m^2$, and large neutrino oscillation
probability (\ref{pro3}); \item large mixing angle citation ???,
$m_2\sim m_1$, i.e., $\Delta m^2\sim 0$, $\theta_m \sim 0$ and
$\Delta m^2_m\simeq A$, leading to small neutrino oscillation
probability (\ref{pro3}).
\end{itemize}
} \comment{ In the framework of SM, neutrinos are massless, $m_1 =
m_2 = 0$ and $\Delta m^2 = 0$, thus they have no mixing and
oscillation between their flavors while traveling in the vacuum.
However, interacting with a normal medium, neutrinos get non-zero
effective masses, and $\Delta m^2_m = A$ [see Eq.~(\ref{media0}],
which is proportional to the electron number density. As a result,
neutrino flavor mixing and oscillation occur while traveling in a
normal medium. }

\section{\large Neutrino scattering in nuclear media}\label{nnint}

\begin{figure}
\centerline{\epsfysize=1.5in\epsfxsize=6in
\epsffile{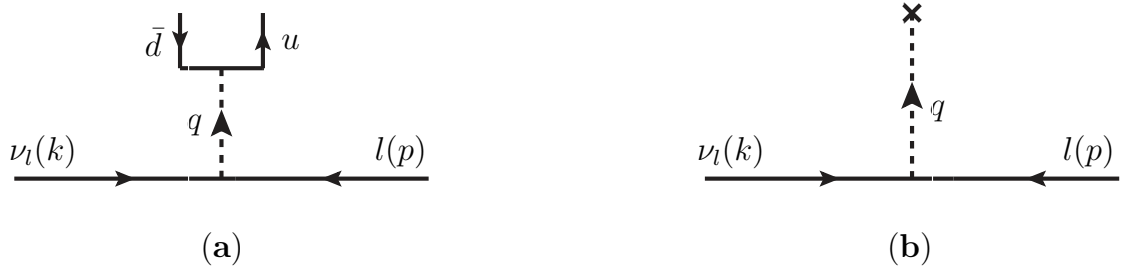}} \caption{
\footnotesize Figure ($a$) indicates an effective interacting vertex (\ref{va}). Figure ($b$) indicates an effective interacting vertex
(\ref{ampn}) of neutrino and lepton scattering with nuclear media represented by ``$\times$'', and the dashed line represents this effective interaction mediated by charged virtual particles
of $W^\pm$ and $\pi^\pm$ fields.
}
\label{vexf}
\end{figure}

In the SM of fundamental particle physics, the effective interacting
Hamiltonian of neutrinos ($\nu_e,\nu_\mu$) interacting with
leptons ($e,\mu$) and quarks ($u,d$) is given by the V-A
current-current interactions (see Fig.~\ref{vexf}-a), that is mediated by a massive charged gauge boson $W^\pm$,
\begin{equation}
{\mathcal H}_w=\frac{G_F}{\sqrt{2}}V_{ud}\Big[\bar
d\gamma^\lambda(1-\gamma_5)u\Big]\Big[\bar
\nu_e\gamma_\lambda(1-\gamma_5)e +\bar
\nu_\mu\gamma_\lambda(1-\gamma_5)\mu\Big]+{\rm h.c.},
\label{va}
\end{equation}
where $G_F$ is the Fermi coupling constant and $V_{ud}$ is the
Cabibbo-Kobayashi-Maskawa (CKM) matrix element. Introducing the
axial current $J^{\,\pi^-}_\lambda$ for a charged pion, one has
\begin{eqnarray}
q_\mu F_\pi &=&\langle 0|J^{\pi^-}_\mu(0)|\pi^-({\bf q})\rangle,
\label{fp}
\end{eqnarray}
where the pion decay constant $F_\pi$ encodes strong interaction effects  \cite{DSM}. The matrix element of the interacting Hamiltonian
(\ref{va}) can be written as an effective interacting vertex in
energy-momentum space
\begin{eqnarray}
{\mathcal V}_{\nu_l,l,\pi}&=&\frac{G_F}{\sqrt{2}}V_{ud}\langle
0|J^{\pi^-}_\lambda(0)|\pi^-({\bf
q})\rangle\bar\nu_l(k)\gamma^\lambda(1-\gamma_5)l(p)+{\rm h.c.}\nonumber\\
&=&\bar G_F q_\lambda[
\bar\nu_l(k)\gamma^\lambda(1-\gamma_5)l(p)]+{\rm h.c.},
\label{ampn}
\end{eqnarray}
where $l=e,\mu$, $\bar G_F\equiv G_FV_{ud}F_\pi/\sqrt{2}$ and the
energy-momentum conservation $q=k+p$. Actually, the vertex
(\ref{ampn}) describes an effective interacting of neutrino
$\bar\nu_l(k)$ and lepton $l^-(p)$ with nuclear matter via virtual pion fields $\pi^-(q)$, as schematically shown in Fig.~\ref{vexf}-b by a dashed line ending with a cross.

\begin{figure}
\centerline{\epsfysize=1.5in\epsfxsize=6in
\epsffile{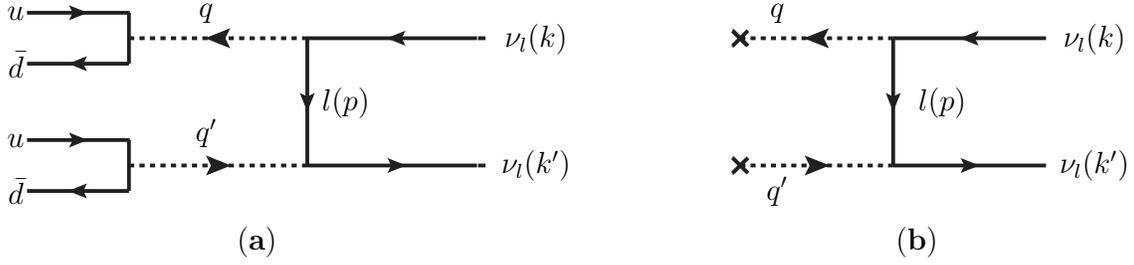}}\caption{ \footnotesize
The Figure ($a$) indicates the effective amplitude
(\ref{ampsq}) of a neutrino scattering with nuclear media ($\times$) by
exchanging two virtual particle fields.
As indicated by Fig.~\ref{vexf}-a, Figure ($b$) indicates the contributions of
the effective interaction (\ref{va}) to the amplitude
(\ref{ampsq}) of neutrino scattering with nuclear media ($\times$). }

\label{ampf}
\end{figure}

Suppose that a neutrino travels through nuclear media, and
interacts with quarks via V-A interactions (\ref{va}), which is
described as an effectively interacting vertex (\ref{ampn}). By
exchanging virtual particles of energy-momentum $q$, a neutrino interacts with nuclear
media and converts into a lepton, which in turn converts back to a
neutrino, as shown in Fig.~(\ref{ampf}). This scattering amplitude
receives all contributions by exchanging even number of virtual
charged particles.  At the leading order (by exchanging two virtual charged particles), we
compute the leading order amplitude ${\mathcal M}_{\nu_l,\nu'_l}$
of neutrino scattering with nuclear media,
\begin{eqnarray}
{\mathcal M}_{\nu_l,\nu'_l}(k,k',q,q',p)
&=&\bar G_F^2
q_\lambda q'_\rho
\bar\nu_l(k)\gamma^\lambda(1-\gamma_5)\big[l(p)\bar l(p)\big]\gamma^\rho(1-\gamma_5)\nu_l(k')\nonumber\\
&=&\bar G_F^2 m_l^2
\bar\nu_l(k)(1+\gamma_5)\big[l(p)\bar
l(p)\big](1-\gamma_5)\nu_l(k'),
\label{ampsq}
\end{eqnarray}
where $q=k+p$, $q'=p+k'$, and in the second line we use Dirac
equations for lepton and massless neutrino, i.e., $\not \!\!p
\,\,l(p)=m_ll(p)$ and $\bar{\nu}(k)\!\!\not\!k \simeq 0$.
Moreover, using the spinor wave-functions $u^{(a)}_l(p)$ of
leptons \cite{zuber}
\begin{eqnarray} [l(p)\bar l(p)]
&=&\sum_{a=1,2}u^{(a)}_l(p)\bar u^{(a)}_l(p)=\frac{\not\!\!p+m_l}{2m_l},
\label{spinor}
\end{eqnarray}
we obtain
\begin{eqnarray}
{\mathcal M}_{\nu_l,\nu'_l}(k,k',q,q',p)
&=&\frac{1}{2}\bar G_F^2 m_l
\bar\nu_l(k)(1+\gamma_5)\not \!\!p(1-\gamma_5)\nu_l(k')\nonumber\\
&=&\frac{1}{2}\bar G_F^2 m_l
\bar\nu_l(k)(1+\gamma_5)(\not \!\!q-\not \!\!k)(1-\gamma_5)\nu_l(k').
\label{ampsq1}
\end{eqnarray}
In order to obtain the total amplitude, we need to integrate over phase spaces of $q$ and $q'$ 
(see Fig.~\ref{ampf}),
\begin{eqnarray}
{\mathcal M}_{\nu_l,\nu'_l}(k,k')
&=&\frac{1}{2}\bar G_F^2 m_l\int_{q}\int_{q'}
\bar\nu_l(k)(1+\gamma_5)(\not \!\!q-\not \!\!k)(1-\gamma_5)\nu_l(k')\nonumber\\
&=&-\frac{1}{2}\bar G_F^2 m_l\int_{q}\int_{q'}
\bar\nu_l(k)(1+\gamma_5)\not \!\!k(1-\gamma_5)\nu_l(k'),
\label{ampsq2}
\end{eqnarray}
where
\begin{eqnarray}
\int_{q}\equiv\int\frac{d^4q}{(2\pi)^4}=\int\frac{d^3q}{2q_0(2\pi)^3},
\label{int2}
\end{eqnarray}
and $q_0\simeq q_0'\approx m_\pi$ pion mass. Because  we study the modification of neutrino energy-spectrum by this scattering, we only consider the case
$k=k'$, which induces $(2\pi)^3\delta^3({\bf q}-{\bf q}')$, as a
result we approximately obtain
\begin{eqnarray}
{\mathcal M}_{\nu_l,\nu'_l}(k,k)
&\approx &-\frac{1}{8}\bar G_F^2\frac{m_l}{m_\pi^2}n_\pi
\bar\nu_l(k)(1+\gamma_5)\not \!\!k(1-\gamma_5)\nu_l(k),
\label{ampsq3}
\end{eqnarray}
where the nucleon number-density
\begin{eqnarray}
n_\pi=\int\frac{d^3q}{(2\pi)^3}.
\label{nde}
\end{eqnarray}

Putting Eq.~(\ref{ampsq3}) together with free kinetic term of
neutrinos propagating through vacuum, we obtain the effective
bilinear Lagrangian for neutrinos,
\begin{eqnarray}
{\mathcal
L}^e_{\bar\nu,\nu} &= &\bar\nu_e(k)(1+\gamma_5)\not\!\! k
\left[1 -{\mathcal A}_e\right](1-\gamma_5)\nu_e(k),
\label{effl}\\
{\mathcal
L}^\mu_{\bar\nu,\nu} &= &\bar\nu_\mu(k)(1+\gamma_5)\not\!\! k
\left[1 -{\mathcal A}_\mu\right](1-\gamma_5)\nu_\mu(k),\label{efflmu}
\end{eqnarray}
where
\begin{equation}
{\mathcal A}_e\approx \left[\frac{G_F}{\sqrt{2}}V_{ud}
F_\pi\right]^2 \left[\frac{m_e}{8m_\pi^2}\right]n_\pi,
\quad {\mathcal A}_{\mu}\approx
\left[\frac{G_F}{\sqrt{2}}V_{ud}F_\pi\right]^2
\left[\frac{m_{\mu}}{8m_\pi^2}\right]n_\pi.
\label{scaa}
\end{equation}
This implies that neutrinos
receive wave-function renormalization when they  propagate through nuclear media.

\comment{
If tau neutrino ($\nu_\tau$) energy is large than the mass of tau lepton $(l=\tau)$, the effective interacting vertex (\ref{ampn}) is also valid for considering tau neutrino scattering with nuclear media.
}
The same discussions and calculations are applied for considering tau neutrino scattering with nuclear media,
\begin{eqnarray}{\mathcal
L}^\tau_{\bar\nu,\nu} &= &\bar\nu_\tau(k)(1+\gamma_5)\not\!\! k
\left[1 -{\mathcal
A}_\tau\right](1-\gamma_5)\nu_\tau(k),\label{effltau}\\
{\mathcal
A}_\tau&\approx& \left[\frac{G_F}{\sqrt{2}}V_{ud}F_\pi\right]^2
\left[\frac{m_{\tau}}{8m_\pi^2}\right]n_\pi.\label{atau}
\end{eqnarray}
We clearly have ${\mathcal A}_\tau\gg {\mathcal A}_\mu\gg
{\mathcal A}_e$. It has to be pointed out that we only consider
interactions of charged currents (\ref{va}) mediated by virtual charged particles, and disregard interactions of neutral
currents. Because the neutral current interaction is universal for all of neutrino and lepton flavors, as will be discussed in
Sec.~\ref{uniform1} for the MSW case.


\section{\large Neutrino oscillations in nuclear media}\label{nuco}

In this section, we turn to study the effects of an uniform
nuclear medium on neutrino oscillations.  Analogously to the MSW approach, using Eqs.~(\ref{h-vac},\ref{effl}), we approximately  obtain the effective
Hamiltonian in the base of neutrino flavor states,
\begin{eqnarray}
\hat{\mathcal H} &=& E+\frac{m_1^2+m_2^2}{4E}+\frac{\Delta
m^2}{4E}
\left(%
\begin{array}{cc}
  -\cos2\theta & \sin2\theta \\
  \sin2\theta & \cos2\theta \\
\end{array}%
\right)+\left(%
\begin{array}{cc}
  E\mathcal A_e & 0 \\
  0 & E\mathcal A_\mu \\
\end{array}%
\right), \label{h-nuc}
\end{eqnarray}
where the last term is due to the contributions of neutrinos scattering with nuclear media.
The same procedure (\ref{h-norm}-\ref{media1}) described in Sec.~\ref{uniform1} leads to the results of the mass squared difference
$\Delta m^2_n$ is,
\begin{eqnarray}
\Delta m^2_n=2\sqrt{(\mathcal A_\mu-\mathcal A_e)^2E^4+\Delta
m^2(\mathcal A_\mu-\mathcal A_e)E^2\cos 2\theta +(\Delta m^2/2)^2}\,,
\label{deltam}
\end{eqnarray}
and the mixing angle is
\begin{eqnarray}
\tan2{\theta}_n&=& \frac{\Delta m^2\sin 2\theta}{\Delta m^2\cos
2\theta+2(\mathcal A_\mu-\mathcal A_e)E^2}\,. \label{newtheta}
\end{eqnarray}
In the base of eigenstates of Hamiltonian (\ref{h-nuc}), neutrino flavor states are represented by
\begin{eqnarray}
|\nu_e\rangle&=&\cos\theta_n|\nu_1\rangle_n+\sin\theta_n|\nu_2\rangle_n\,,
\nonumber\\|\nu_\mu\rangle&=&-\sin\theta_n|\nu_1\rangle_n+\cos\theta_n|\nu_2\rangle_n\, .
\label{newbasen}
\end{eqnarray}

Based on these results (\ref{deltam},\ref{newtheta}) and using Eq.~(\ref{pro2}), we obtain the probability
of neutrino oscillations in nuclear medium
\begin{eqnarray}
P^n_{\nu_{\mu}\leftrightarrow\nu_e}(t)&=&\sin^22\theta_n\sin^2\left[(E^n_2-E^n_1)t
\right],
\label{pro4}
\end{eqnarray}
where
\begin{eqnarray}
{E}^n_2-{E}^n_1&=&\frac{\Delta
m^2_n}{2{E}}.
\label{ed-nuc}
\end{eqnarray}
When $\mathcal
A_e=\mathcal A_\mu=0$, it reduces to neutrino oscillation in the vacuum.

From Eq.~(\ref{scaa}) for the Fermi coupling constant
$G_F=1.16\times 10^{-5}\,\rm{GeV}^{-2}$, the pion decay constant
$F_\pi=130.41\,\,\rm MeV$ \cite{data group}, $m_\mu/m_\pi\simeq
0.5$ and $n_\pi\sim 10^{38}\,{\rm cm}^{-3}$ in nuclear media, we
have
\begin{eqnarray}
\mathcal A_e\simeq 3\times10^{-16},\quad \mathcal A_\mu\simeq 3\times10^{-14}.
\label{a}
\end{eqnarray}
In this case we approximately have
\begin{eqnarray}
\Delta m^2_n &\approx&  2\sqrt{\mathcal A_\mu^2E^4+\Delta
m^2\mathcal A_\mu E^2\cos 2\theta+(\Delta m^2/2)^2}\,,
\label{adeltam}\\
\tan2{\theta}_n &\approx& \frac{\Delta m^2\sin 2\theta}{\Delta m^2\cos
2\theta+2\mathcal A_\mu E^2}\,.
\label{anewtheta}
\end{eqnarray}
We discuss the following two cases. The case (i) small vacuum
mixing angle $\theta \sim 0$ and from Eq.~(\ref{ed-nuc}), we
approximately have
\begin{eqnarray}
{E}^n_2-{E}^n_1&\simeq&\frac{\Delta
m^2}{2E}+\mathcal A_\mu E,\quad \tan 2\theta_n\approx 0. \label{ed-nuc1}
\end{eqnarray}
The case (ii) large vacuum mixing angle $\theta \sim \pi/4$ and
Eq.~(\ref{ed-nuc}) is approximated to be
\begin{eqnarray}
{E}^n_2-{E}^n_1&\simeq& \sqrt{\left(\frac{\Delta m^2}{2E}\right)^2 +({\mathcal A}_\mu E)^2},\quad \tan 2\theta_n\approx \frac{\Delta m^2}{2{\mathcal A}_\mu E^2}.
\label{ed-nuc2}
\end{eqnarray}
The first case implies that the nuclear medium effect could possibly be relevant for neutrino oscillations when neutrino energy is very large. In the second case for large neutrino energy, however, the mass squared difference is dominated by the nuclear effect $({\mathcal A}_\mu E)^2$ and mixing angle $\theta_n$ becomes smaller.

\begin{figure}
  \includegraphics[width=4.9in]{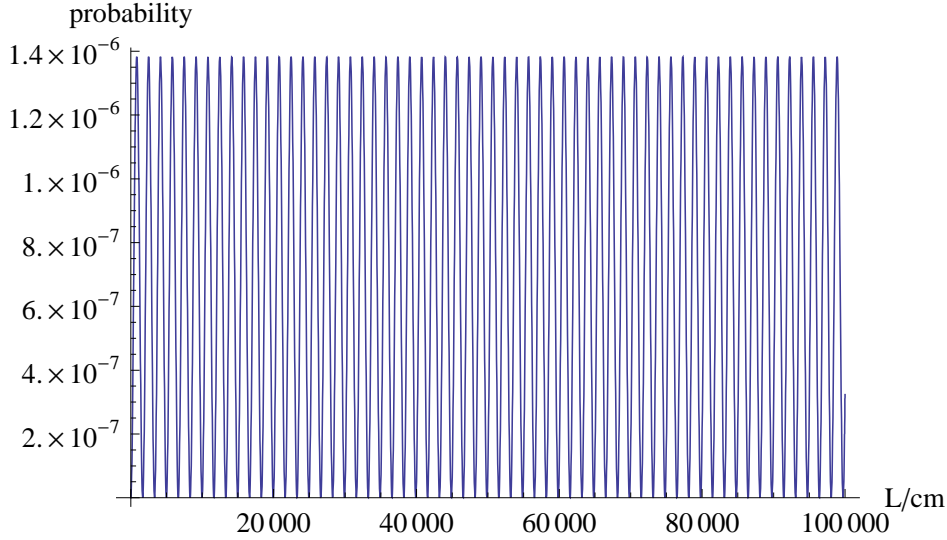}\\
  \caption{The probability (\ref{pro4}) of electron neutrinos converting to muon neutrinos in the presence of nuclear medium is plotted as a function of traveling distance $L\simeq c t$. Neutrino energy $E=1$MeV.}
\label{n_o_n}
\end{figure}

\begin{figure}
  \includegraphics[width=4.9in]{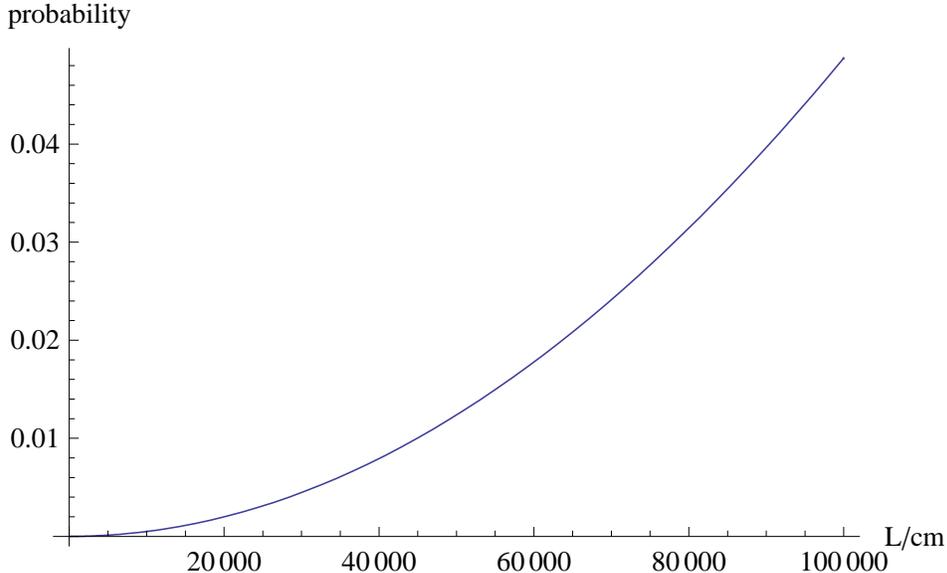}\\
  \caption{The probability (\ref{pro4}) of electron neutrinos converting to muon neutrinos in the absence of nuclear medium (the vacuum oscillation) is plotted as a function of traveling distance $L\simeq c t$. Neutrino energy $E=1$MeV.}
\label{n_o_v}
\end{figure}

In order to see the nuclear medium effect on neutrino oscillations, we consider the case of neutron stars or quark stars. Suppose that electron neutrinos $\nu_e$ are produced in the core of neutron stars and their average energy is in the $1{\rm MeV}\lesssim \bar E\lesssim 10$MeV \cite{rews1}. Based on Eqs.~(\ref{ed-nuc}-\ref{anewtheta}) for neutrino neutrino energy $E=1$MeV, we calculate the probability (\ref{pro4}) of electron neutrinos converting to muon neutrinos when they travel from the center to surface of stars. In calculations, we use the vacuum oscillation parameters: the squared
mass difference $\Delta m^2_{\rm sun}\simeq 7.59\times
10^{-5}{\rm eV}^2$ and $\tan^2\theta_{\rm sun} \simeq 0.47$ from the
Solar neutrino experiment \cite{exp-solar}. The result is plotted in Fig.~\ref{n_o_n}. As a comparison, we make the same calculation in absence of nuclear matter (the vacuum oscillation) and result is plotted in Fig.~\ref{n_o_v}. These plots show that in the absence of nuclear medium (vacuum oscillation), the probability of electron neutrino converting to muon neutrino at the surface of stars is about $5\%$. While in the presence of nuclear medium, the probability of electron neutrino converting to muon neutrino at the surface of stars is about $7\times 10^{-7}$, averaged over traveling distance. This means that most electron neutrinos remain electron neutrinos. By this compariosn, it implies that the effect of nuclear medium on neutrino oscillations should be considered when
one studies properties of neutrinos created inside neutron stars and quark stars, in particular for high-energy neutrinos. When neutrino energy $E\sim 0.1$MeV, $E{\mathcal A}_\mu \sim (\delta m^2/2E)$, the effect of nuclear media $E{\mathcal A}_\mu$ is comparable with $(\delta m^2/2E)$ [see Eqs.~(\ref{a},\ref{ed-nuc2})].  Therefore low-energy neutrinos $E< 0.1$MeV, the effect of nuclear media on neutrino oscillations is not important, and neutrino oscillation pattern is slightly deviated from the pattern of vacuum neutrino oscillations.      

\comment{
Both cases imply that the nuclear medium effect could possibly be relevant when
\begin{eqnarray}
m_{2,1}^2 \gtrsim 2 \Delta m^2 {\mathcal A}^{-1}_\mu \simeq 6\times 10^{13} \Delta m^2 .
\label{con1}
\end{eqnarray}
In this case, assuming $m_2\simeq m_1\simeq 10
\,\,\rm{ KeV}$, we have the nuclear medium effect
\begin{eqnarray}
\Delta m^2\simeq \frac{1}{2}{\mathcal{A}}_\mu m^2_2 \simeq
6\times 10^{-6}\rm{eV^2} \,\,.
\end{eqnarray}
This is consistent with the experimental evidences of the squared
mass difference $\Delta m^2_{\rm sun}\simeq 7.59\times
10^{-5}eV^2$ and $\tan^2\theta_{\rm sun} \simeq 0.47$ for the
Solar neutrino experiment \cite{exp-solar}.  
and
$\Delta m^2_{\rm atom}\simeq 2.43\times 10^{-3}eV^2$ and
$\tan^2\theta_{\rm atom} > 90\%$ for atmospheric neutrinos
\cite{exp-atm}.
}


\section{\large Neutrino oscillations in strongly magnetized nuclear
media} \label{mag}

In the SM for particle physics, if neutrinos are massive, they
have electric dipole and magnetic moment due to quantum
corrections. Therefore neutrinos interact with electromagnetic
fields, although they are electrically neutral. \comment{ Neutrino
electromagnetic properties are important because they are directly
connected to fundamentals of particle physics and can be used to
distinguish Dirac and Majorana neutrinos. } Neutrino magnetic
moment is given by
\begin{eqnarray}
\mu_i&=&\frac{3G_Fe}{8\sqrt{2}\pi^2}m_i,\quad i=1,2, \label{mu}
\end{eqnarray}
which is proportional to neutrino masses $m_i$  \cite{lee}. In the unit of the
Bohr magneton $\mu_B=e/(2m_e)$,
$\mu_i=3.1\times10^{-19}\mu_B(m_i/1eV)$. The interacting Hamiltonian is diagonal in the base of mass eigenstates and given by
\begin{eqnarray}
\mathcal H_b
=-\mu_iB=-\frac{3G_Fm_e^2}{8\sqrt{2}\pi^2}\left(\frac{B}{B_c}\right)m_i,
\label{intmu}
\end{eqnarray}
where the critical field $B_c\equiv m_e^2/e= 4.3\times 10^{13}$
Gauss. The interacting Hamiltonian (\ref{intmu}) is very small,
$\mathcal H^b\sim 10^{-2}(m_e/m_W)^2m_i \ll m_i$ for $B=B_c$,
where $m_W\simeq 80$GeV is the $W$-gauge boson mass \cite{data
group}. Therefore this interaction could be relevant only for very
strong magnetic fields $B$.

Using Eqs.~(\ref{h-nuc}) and (\ref{mu}), the effective Hamiltonian
in the base of flavor eigenstates is given by
\begin{eqnarray}
\hat{\mathcal H}_{nb} \!\!&=& \!\!\hat{\mathcal H}_n+ \hat{\mathcal H}_b\nonumber\\
&=& \!\!E+\frac{M_1+M_2}{2}+
\!\!\frac{\Delta M^2}{4E}\left(%
\begin{array}{cc}
  -\cos2\theta &  \sin2\theta \\
  \sin2\theta &   \cos2\theta\\
\end{array}%
\!\!\right)+\left(%
\begin{array}{cc}
  E\mathcal A_e & 0 \\
  0 & E\mathcal A_\mu \\
\end{array}%
\right),
 \label{nucmag}
\end{eqnarray}
where
\begin{eqnarray}
M_i\equiv \frac{m^2_i}{2E}-\mu_iB, \quad \frac{\Delta
M^2}{2E}\equiv \frac{\Delta m^2}{2E} -(\mu_2-\mu_1)B. \label{bm}
\end{eqnarray}
Following the same calculations (\ref{deltam}-\ref{pro4}), we
obtain the probability of neutrino oscillations in strongly
magnetized nuclear media
\begin{eqnarray}
P^{nb}_{\nu_{\mu}\leftrightarrow\nu_e}(t)&=&\sin^22\theta_{nb}\sin^2\left[(E^{nb}_2-E^{nb}_1)t
\right],
\label{pro5}
\end{eqnarray}
where the mass squared difference
is given by,
\begin{eqnarray}
\Delta m^2_{nb}=2\sqrt{(\mathcal A_\mu-\mathcal A_e)^2E^4+\Delta
M^2(\mathcal A_\mu-\mathcal A_e)E^2\cos 2\theta +(\Delta
M^2/2)^2}\,, \label{deltamb}
\end{eqnarray}
and the mixing angle is
\begin{eqnarray}
\tan2{\theta}_{nb}&=& \frac{\Delta M^2\sin 2\theta}{\Delta M^2\cos
2\theta+2(\mathcal A_\mu-\mathcal A_e)E^2}\,. \label{newthetab}
\end{eqnarray}

Analogously to the previous section, considering the case (i) small vacuum mixing angle
$\theta \sim 0$ we approximately have
\begin{eqnarray}
{E}^{nb}_2-{E}^{nb}_1&\simeq&\frac{\Delta m^2}{2E}-(\mu_2-\mu_1)B
+ \mathcal A_\mu E,\quad \tan 2\theta_{nb}\approx 0.
\label{ed-nuc1b}
\end{eqnarray}
Whereas the case (ii) large vacuum mixing angle $\theta \sim \pi/4$ and
Eq.~(\ref{ed-nuc}) is approximated to be
\begin{eqnarray}
{E}^{nb}_2-{E}^{nb}_1&\simeq& \sqrt{\left[\frac{\Delta
m^2}{2E}-(\mu_2-\mu_1)B\right]^2 +({\mathcal A}_\mu E)^2},\nonumber\\
\tan 2\theta_{nb} &\approx & \frac{\frac{\Delta
m^2}{2E}-(\mu_2-\mu_1)B}{{\mathcal A}_\mu E}. \label{ed-nuc2b}
\end{eqnarray}
The first case implies that the nuclear medium effect could
possibly be relevant for neutrino oscillations when neutrino
energy is very large.

\comment{
\begin{eqnarray}
\Delta m^2_{nb}&=&2\Big[X^2E^2+\Delta m^2XE\cos
2\theta +(\frac{\Delta m^2}{2})^2\nonumber\\
&-&Y^2E^2+\Delta m^2YE\sin 2\theta\Big]^{1/2}, \label{deltamb}
\end{eqnarray}
where
\begin{eqnarray}
X&=&E(\mathcal A_\mu-\mathcal A_e)-(\mu_2-\mu_1)B\cos 2\theta,\nonumber\\
Y&=& (\mu_2-\mu_1)B\sin2\theta,\label{xyb}
\end{eqnarray}
and the mixing angle
\begin{eqnarray}
\tan2{\theta}_{nb}&=& \frac{[\Delta
m^2-2(\mu_2-\mu_1)BE]\sin2\theta}{[\Delta
m^2-2(\mu_2-\mu_1)BE]\cos 2\theta+2(\mathcal A_\mu-\mathcal
A_e)E^2}\,. \label{newthetab}
\end{eqnarray}
The neutrinos energy difference is
\begin{eqnarray}
{E}^{nb}_2-{E}^{nb}_1&=&\frac{\Delta m^2_{nb}}{2E}
\,, \label{ed-nuc}
\end{eqnarray}
and $\mathcal A_e$-term is neglected as $\mathcal A_e \ll \mathcal
A_\mu$. In the absence of nuclear media $\mathcal A_e =\mathcal
A_\mu =0$, above equations determine the probability of neutrino
oscillations in strong magnetic fields.
}
\section{Neutrino oscillations in general case}\label{g}

In this sections we give a general discussion of neutrino
oscillations in neutral nuclear media with the presence of both electrons
and strong magnetic fields. If electron presents in nuclear media,
the $V_C$-term (\ref{vc}) should be added into the component
$\hat{\mathcal H}_{11}$, as  discussed in Sec.~\ref{uniform1}. For
this general case the Hamiltonian in the base of neutrino flavor eigenstates is given
by
\begin{eqnarray}
 \hat{\mathcal{H}}_{nmb}&=& \hat{\mathcal{H}}_{nb}+\hat{\mathcal{H}}_{m}\nonumber\\
&=&\!\!E+\frac{M_1+M_2}{2}-\frac{1}{\sqrt 2} G_F
n_n+\frac{\Delta M^2}{4E}\left(%
\begin{array}{cc}
  -\cos2\theta &  \sin2\theta \\
  \sin2\theta &   \cos2\theta\\
\end{array}%
\!\!\right)\nonumber\\
\!\!&+&\left(%
\begin{array}{cc}
  E\mathcal A_e+\sqrt 2 G_F n_e & 0 \\
  0 & E\mathcal A_\mu \\
\end{array}%
\right) . \label{nmb}
\end{eqnarray}
Diagonalizing Eq.~(\ref{nmb}), we obtain the eigenvalues of
$\hat{\mathcal{H}}_{nmb}$ in mass eigenstates
\begin{eqnarray}
\hat E_i&=&E-\frac{1}{\sqrt{2}}G_Fn_n+\frac{\tilde{m}^2_{i}}{2E},
\label{enmb}
\end{eqnarray}
where $\tilde{m}_i$ are effective neutrino masses in the presence
of electrons, nuclear media and magnetic fields. As a result, the
effective neutrino mass squared difference and mixing angles are
given by
where the mass squared difference
is given by,
\begin{eqnarray}
\Delta \tilde m^2=2\sqrt{(\mathcal A_\mu - A_e)^2E^4+\Delta
M^2(\mathcal A_\mu-A_e)E^2\cos 2\theta+(\Delta M^2/2)^2}\,,
\label{mnmb}
\end{eqnarray}
and the mixing angle is
\begin{eqnarray}
\tan2{\tilde \theta}&=& \frac{\Delta M^2\sin 2\theta}{\Delta M^2\cos
2\theta+2(\mathcal A_\mu-A_e)E^2}\,,
\label{tnmb}
\end{eqnarray}
where $A_e\equiv \mathcal A_e+\sqrt 2 G_F n_e/E$.

\comment{
\begin{eqnarray}
\Delta\tilde{m}^2&=& 2\Big[(X_1+X_3)^2E^2+\Delta m^2(X_2-X_3)E\cos
2\theta -4E^2X_2X_3+(\frac{\Delta m^2}{2})^2\nonumber\\
&-&4Y^2E^2-2\Delta m^2YE\sin 2\theta\Big]^{1/2},\label{mnmb}
\\
\tan2{\tilde\theta}&=& \frac{[\Delta
m^2-2(\mu_2-\mu_1)BE]\sin2\theta}{(\Delta
m^2-2(\mu_2-\mu_1)BE)\cos 2\theta+2(\mathcal A_\mu-\mathcal
A_e)E^2-A}\,,\label{tnmb}
\end{eqnarray}
where $X_3=E\mathcal A_e-\mu_1B\cos^2\theta-\mu_2B\sin^2\theta-A$.
}
Neutrino flavor states are expressed by
\begin{eqnarray}
|\nu_e\rangle&=&\cos\tilde\theta|\tilde\nu_1\rangle+\sin\tilde\theta|\tilde\nu_2\rangle\,,
\nonumber\\|\nu_\mu\rangle&=&-\sin\tilde\theta|\tilde\nu_1\rangle+\cos\tilde\theta|\tilde\nu_2\rangle\, ,
\label{newbaseng}
\end{eqnarray}
in terms of eigenstates $|\tilde\nu_i\rangle$ of Hamiltonian (\ref{nmb}).
The probability of neutrino oscillations is given
by
\begin{eqnarray}
P^{nmb}_{\nu_{\mu}\leftrightarrow\nu_e}(t)&=&\sin^22\tilde\theta\sin^2\left[(E^{nmb}_2-E^{nmb}_1)t
\right],
\label{pro6}
\end{eqnarray}
where the vacuum mixing angle $\theta$ is changed to $\tilde \theta$
(\ref{tnmb}), and the neutrino energy difference is changed to
\begin{eqnarray}
{E}^{nmb}_2-{E}^{nmb}_1&=&\frac{\Delta \tilde m^2}{2E}.
\label{nmbe}
\end{eqnarray}
In the absence of nuclear media and magnetic fields,
Eqs.~(\ref{pro6}) and (\ref{nmbe}) reduces to the MSW case
Eqs.~(\ref{pro3}) and (\ref{em12}).

Neglecting the term $\mathcal A_e$ [see Eq.~(\ref{a}] and the effect of magnetic fields,we approximately have
\begin{eqnarray}
\Delta \tilde m^2 &\approx&  2\sqrt{\chi^2E^4+\Delta
m^2\chi E^2\cos 2\theta+(\Delta m^2/2)^2}\,,
\label{adeltam2}\\
\tan2{\tilde\theta}&\approx& \frac{\Delta m^2\sin 2\theta}{\Delta m^2\cos
2\theta+2\chi E^2},
\label{anewtheta2}\\
\chi &\equiv& \mathcal A_\mu -\sqrt 2 G_F n_e/E.
\label{cnmsw}
\end{eqnarray}
Assuming that in neutral nuclear media the electron density is about three orders of magnitude smaller than nuclear density $n_e\sim 10^{-3} n_\pi$, as the case of neutron stars, in Eq.~(\ref{cnmsw}), we compare the nuclear effect $\mathcal A_\mu$ [see Eq.~(\ref{a}] with the MSW effect (\ref{vc}),
\begin{eqnarray}
\frac{\sqrt 2 G_F n_e}{E} \sim 10^{-9}\frac{m_\pi}{E},
\label{compmn}
\end{eqnarray}
both effects are comparable for very high neutrino energy $E\sim
10^5\times m_\pi $. However, if neutral nuclear media have very
small densities of electrons and protons, the nuclear effect
($\mathcal A_\mu$) is more important than the MSW effect.

\section{Neutrino oscillation resonance and level-crossing
} \label{reson}

In this section, we turn to consider effects of non-uniform media
on neutrino oscillations. As discussed neutrino oscillation in
media depends on electron and baryon densities. Thus the variation of these
densities changes the oscillation parameters, mass squared
difference (\ref{media0},\ref{deltam}) and mixing angle
(\ref{media1},\ref{newtheta}). Therefore, on the way of neutrino
traveling through non-uniform media, the probability of neutrino
oscillations is changing and reaches a resonance at which
neutrino oscillation is maximal even if the vacuum mixing angle is small.

\subsection{MSW resonance and level-crossing
}\label{mswr}

First we recall discussions on neutrino oscillations in a normal
medium and resonance of maximal neutrino mixing and oscillations
\cite{wolf,smirniv}. When neutrinos propagate through media
where electron density $n_e$ varies, in consequence the
probability of neutrino oscillations changes, as the mass squared
difference $\Delta m^2_m$ (\ref{media0}) and mixing angle
$\theta_m$ (\ref{media1}) vary. Then the flavor composition of the
neutrinos $(\nu_e,\nu_\mu)$ along their traveling path is a function of the
electron density profile, and the phenomena of the MSW resonance
and level-crossing may occur.

For simplicity, we discuss these phenomena for a small vacuum
mixing angle ($\theta\ll 1$). In this case, Eq.~(\ref{newbase})
shows that the electron neutrino $|\nu_{e}\rangle$ is mainly in
the mass eigenstate $|\nu_1\rangle$ and the muon neutrino
$|\nu_{\mu}\rangle$ is mainly $|\nu_2\rangle$. If the electron
density is so small $n_e\simeq 0$ that $2\sqrt{2}G_Fn_eE\ll \Delta
m^2 \cos 2\theta$ in Eqs.~(\ref{media0},\ref{media1}), then we
have $\theta_m\simeq\theta\sim 0$, 
and
\begin{eqnarray}
|\nu_1\rangle_m \simeq |\nu_1\rangle, \quad |\nu_2\rangle_m \simeq
|\nu_2\rangle \,\,.
\label{a=0}
\end{eqnarray}
Eq.~(\ref{newbasem}) tells us that the electron neutrino
$|\nu_{e}\rangle$ is mainly in the state $|\nu_1\rangle_m$ and the
muon neutrino is $|\nu_{\mu}\rangle$ mainly $|\nu_2\rangle_m$,
i.e., the flavor composition $(\nu_e,\nu_\mu)$ of the neutrinos is
\begin{eqnarray}
|\nu_e\rangle\approx |\nu_1\rangle_m, \quad |\nu_\mu\rangle \approx |\nu_2\rangle_m\,.
\label{noflip}
\end{eqnarray}
If the electron density is a such critical value
\begin{eqnarray}
n^c_e=\frac{\Delta m^2}{\sqrt{2}G_F E} \cos 2\theta\,,
\label{crin}
\end{eqnarray}
that $2\sqrt{2}G_Fn_eE=\Delta m^2 \cos
2\theta$, Eq.~(\ref{media1}) gives rise a maximal mixing angle
$\theta_m\approx \pi/4$ 
, as a result,
maximal probability of the neutrino oscillation occurs, {\it the
MSW resonance}. The flavor composition $(\nu_e,\nu_\mu)$ of the
neutrinos is
\begin{eqnarray}
|\nu_e\rangle&=&\frac{1}{\sqrt{2}}\big(|\nu_1\rangle_m+|\nu_2\rangle_m\big)\,,\nonumber\\
|\nu_\mu\rangle&=&\frac{1}{\sqrt{2}}\big(|\nu_2\rangle_m
-|\nu_1\rangle_m\big)\, .
\label{newbasem1}
\end{eqnarray}

If the electron density is so large that $2\sqrt{2}G_Fn_eE\gg \Delta m^2 \cos
2\theta$, i.e., $n_e\gg n_e^c$, Eq.~(\ref{media1}) shows that
$\tan 2\theta_m \rightarrow 0^-$, $\theta_m\rightarrow \pi/2$ 
. Eq.~(\ref{newbasem}) gives the
flavor composition
\begin{eqnarray}
|\nu_e\rangle\approx |\nu_2\rangle_m, \quad |\nu_\mu\rangle \approx -|\nu_1\rangle_m\,.
\label{flip}
\end{eqnarray}
Comparison between the flavor composition (\ref{noflip}) for small electron
density  with the one (\ref{flip})
for large electron density  shows
an inversion of neutrino flavors. This neutrino flavor inversion
is known as {\it the MSW level-crossing} of neutrino flavors.

\subsection{\large Resonance and level-crossing in the general case}\label{leveln}

As discussed in the previous section 
for the MSW
resonance, when the dominator of Eq.~(\ref{tnmb}) vanishes
\begin{eqnarray}
[\Delta m^2-2(\mu_2-\mu_1)BE]\cos 2\theta+2(\mathcal
A_\mu-\mathcal A_e)E^2-2\sqrt 2G_Fn_eE=0\,, \label{reg0}
\end{eqnarray}
the effective neutrino mixing angle is maximal ($\tilde \theta
\sim \pi/4$), leading to a resonance of the neutrino oscillation
probability in the general case. The condition (\ref{reg0}) for
the resonance gives a critical electron density
\begin{eqnarray}
n^c_e &\simeq&\left(\frac{1}{\sqrt
2G_F}\right)\Big[\left(\frac{\Delta
m^2}{2E}-(\mu_2-\mu_1)B\right)\cos 2\theta+(\mathcal
A_\mu-\mathcal A_e)E\Big], \label{ed-nucg}
\end{eqnarray}
as a function of neutrino energy, nuclear density $n_\pi$ and
magnetic field $B$, in addition to vacuum mixing angle $\theta$,
and neutrino masses $m_{1,2}$. Note that the first term in
Eq.~(\ref{ed-nucg}) corresponds the MSW critical density, and
other terms come from the effects of magnetized nuclear media.
When $n_\pi=0$ and $B=0$, critical density (\ref{ed-nucg}) becomes
to the MSW one (\ref{crin}).

Analogously to the MSW level-crossing, we discuss the
level-crossing of neutrino oscillations in the general case.
Suppose that the vacuum mixing angle is small ($\theta\ll 1$),
magnetic fields, nuclear media and electrons densities are small,
i.e., $n_\pi\sim 0,\,\,n_e\sim 0,\, B\sim 0$. Then, from
Eqs.~(\ref{tnmb},\ref{newbaseng}) $\tilde{\theta}\simeq\theta\sim
0$ 
, which implies that the electron neutrino $|\nu_{e}\rangle$ is
mainly in the mass eigenstate
$|\tilde{\nu}_1\rangle\simeq|\nu_1\rangle $ and the muon neutrino
$|\nu_{\mu}\rangle$ is mainly
$|\tilde{\nu}_2\rangle\simeq|\nu_2\rangle $, similarly to
Eq.~(\ref{a=0}). As increasing magnetic field, nucleon and
electron densities, from Eq.~(\ref{tnmb}), the neutrino mixing
angle increases, and the resonance takes place if the condition
(\ref{reg0}) is satisfied. In this case, Eqs.~(\ref{mnmb}) and
(\ref{tnmb}) give rise a maximal mixing $\tilde\theta\approx
\pi/4$ 
, and maximal probability of neutrino oscillations occurs. The
flavor composition $(\nu_e,\nu_\mu)$ of the neutrinos is
\begin{eqnarray}
|\nu_e\rangle&=&\frac{1}{\sqrt{2}}\big(|\tilde\nu_1\rangle+|\tilde\nu_2\rangle\big)\,,\nonumber\\
|\nu_\mu\rangle&=&\frac{1}{\sqrt{2}}\big(|\tilde\nu_2\rangle
-|\tilde\nu_1\rangle\big)\, .
\label{newbasem2}
\end{eqnarray}
As further increasing magnetic field $B$, electron and nucleon
densities, from Eqs.~(\ref{tnmb},\ref{reg0}), the condition for
the MSW level-crossing to occur is that
\begin{eqnarray}
[\Delta m^2-2(\mu_2-\mu_1)BE]\cos 2\theta+2(\mathcal
A_\mu-\mathcal A_e)E^2-2\sqrt 2G_Fn_eE \rightarrow \infty^-,
\label{flipng}
\end{eqnarray}
i.e., $\tan 2\tilde\theta\rightarrow 0^-$,
$\tilde\theta\rightarrow \pi/2$ 
, which is the same as the MSW result. The flavor
composition is
\begin{eqnarray}
|\nu_e\rangle\approx |\tilde\nu_2\rangle, \quad |\nu_\mu\rangle \approx -|\tilde\nu_1\rangle\,.
\label{flipn}
\end{eqnarray}
This shows an inversion of neutrino flavors, {\it the MSW
level-crossing} of neutrino flavors, as discussed in the previous
section.

\section{\large Summary and remarks} \label{conc}

Neutrino mass, mixing and oscillation are fundamental issues to be
completely understood beyond the standard model of elementary
particle physics. It still needs much effort to understand the
origin of neutrino masses ($m_1, m_2, m_3$) and mixing angles
($\theta_{12},\theta_{23},\theta_{13}$) and $CP$ phase $\delta$,
and so far they can be treated as fundamental parameters, which
are probably not independent \cite{xue}. In this article,
we study the mixing and
oscillation of electron and muon neutrinos, base on the assumption that the mixing angle $\theta_{12}$ between the first
and second generations is much larger than the mixing angles $\theta_{23}$ and $\theta_{13}$.

After a brief review of vacuum neutrino oscillations and matter
neutrino oscillations (MSW solution), we study neutrino
oscillations in strongly magnetized nuclear media. For this purpose, we
first compute the effective Hamiltonian of neutrino interacting
with nuclear media up to the leading order $\mathcal O(G_F^2)$. As shown in Fig.~\ref{ampf}, this is 
in fact the process of neutrino absorption ($\nu + n\rightarrow p +e^-$) then neutrino emission ($p + e^-\rightarrow n + \nu$). The leading order contributions to both neutrino absorption and emission
are $\mathcal O(G_F)$. In order to obtain the neutrino energy spectrum in nuclear media, we set neutrino incoming momentum ``$k$'' to be the same as neutrino outgoing momentum ``$k'$'' in Eq.~(\ref{ampsq2}), see Fig.~\ref{ampf},
based on the assumption that nuclear medium is a very massive and rigid solid so that the recoiling effect is completely negligible, ${\bf q}=-{\bf q}'$ in Eq.~(\ref{ampsq2}). According to Eqs.~(\ref{fp}) and (\ref{ampn}), this assumption can also be viewed as a scenario that a propagating neutrino ``$\nu_{e,\mu,\tau}(k)$'' interacts with short ranged fields $W^+(q)$ and $\pi^+(q)$ in nuclear media, as a result its wave function receives the correction (${\mathcal A}_{e,\mu,\tau}$). This approximation should be further examined to see the effect of ``incoherent'' neutrino scattering with nuclear media for $k\not= k'$ on the neutrino oscillation in nuclear media that we consider.  With this
Hamiltonian, we calculate the effective mixing angle and mass squared
difference, as a result we obtain the probability of neutrino
oscillations in nuclear media and strong magnetic fields. Moreover, we
discuss the resonance and level-crossing of neutrino oscillations
in magnetized nuclear media. It is shown that the effects due to
magnetized nuclear media can modifies the pattern of vacuum
neutrino oscillations. As example, we calculate the effect of nuclear media on neutrino oscillation, and find that neutrino oscillations are suppressed for high-energy neutrino $E>1$MeV. These effects on neutrino oscillations
need to be considered in studying neutrino physics in supernovae,
neutron stars, quark stars and magnetars, where neutrinos produced are very energetic $1{\rm MeV}\lesssim E\lesssim 10{\rm MeV}$ \cite{rews1}. It is also interesting to see whether
or not the neutrino oscillation in nuclear media has any effect on
the decay of heavy ions  \cite{Kleinert_2009}.

Moreover, it is interesting to study nuclear medium effects on
neutrino mixing and oscillation in the three generation case,
electron, muon and tau neutrinos $(\nu_e,\nu_\mu,\nu_\tau)$
(\ref{ij}-\ref{prob}) to find modifications of the vacuum mixing
angles $\theta_{ij}$ and mass squared difference $\Delta
m^2_{ij}$, and the $CP$ phase $\delta$ \cite{xueimanprep}. In this
case, the effect of tau neutrino scattering with nuclear media
[see Eq.~(\ref{effltau},\ref{atau})] should be important.

\section*{\small Acknowledgment}

We thank to E.~Bavarsad
for discussions. One of authors, S.-S.~Xue thanks Professor Hagen Kleinert for many discussions,
and I.~Motie thanks Professor R.~Ruffini for his
hospitality, when he stays in ICRANet Pescara, Italy, where this
work is done.


\begin{thebibliography}{99}

\bibitem{pontecrov}
B.~Pontecorvo, Journal of Experimental and Theoretical Physics, 6
429 (1957); V.~Gribov, B.~Pontecorvo, Phys.~Lett.~B 28 493 (1969).

\bibitem{exp-solar}
B.~T.~Cleveland et al. Astrophysics. J. 496 (1998); S.~N.~Ahmed et
al., Phys. Rev. Lett. 92 181301 (2004).

\bibitem{exp-reactor}
K.~Eguchi et al., Phys. Rev. Lett. 90 021802 (2003);\\
K.~Eguchi et
al., Phys. Rev. Lett. 94 081801 (2005).

\bibitem{exp-atm}
S~.H~.~Hirata et al. Phys. Lett. B 280 146(1992); Y.~Fukuda et
al., Phys. Lett. B 335 237 (1994); S.~Fukuda et al., Phys. Rev.
Lett. 81
 1562 (1998); Y.~Ashie et al., Phys.~Rev.~Lett. 93, 101801 (2004).

\bibitem{exp-accel}
M.~H.~Ahn et al. Phys. Rev. Lett. 90 041801 (2003).

\bibitem{wolf}
L.~Wolfenstein, Phys. Rev. D 17, 2369 (1978).

\bibitem{smirniv}
S.~P.~Mikheyev, A.~Y.~Smirnov, NuCim C9 17 (1986).

\bibitem{mohapatra}
R.~N.~Mohapatra and P.~B.~Pal, Massive Neutrinos in Physics and
Astrophysics, Singapore: World Scientific Publishing Co. Pte. Ltd.
(2003).

\bibitem{funda}
C.~Giunti, C.~W.~Kim, Fundamentals of Neutrino Physics and
Astrophysics, New York: Oxford University Press Inc, (2007).

\bibitem{zober}
K.~Zuber, Neutrino Physics, New York: Taylor and Francis Group,
LLC, (2004).

\bibitem{DSM} See for example,
J.~F.~Donoghue, E.~Golowich and B.~R.~Holstein, Dynamics of the
standard model, New York: Cambridge University Press , (1992);
T.~Morii, C.~S.~Lim and S.~N.~Mukherjee, The Physics of the
Standard Model and Beyond, Singapore: World Scientific Publishing
Co. Pte. Ltd. (2004).

\bibitem{zuber}
C.~Itzykson, J.~B.~Zuber: Quantum field theory, McGraw-Hill:
United States of America (1980).

\bibitem{data group}
W.-M. Yao et al., Particle Data Group, J. Phys. G 33, 1 (2006).

\bibitem{rews1} see for example, D.~Arnett, 1996, ``Supernovae and Nucleosynthesis'', Princeton
University Press, Princeton NJ.

\bibitem{lee}
B.~W.~Lee and R.~E.Shrock, Phys.~Rev.~D 16, 1444 (1977).

\bibitem{xue}
She-Sheng Xue, Mod.~Phys.~Lett.~A 14, 2701 (1999)
[arXiv:hep-ph/9706301v1].

\bibitem{Kleinert_2009}
H.~Kleinert and P.~Kienle, EJTP 6, No. 22, 107–122 (2009),
[arXiv:nucl-th/0803.2938v4], references therein.

\bibitem{xueimanprep}
I.~ Motie and S.-S.~Xue, in preparation.

\end{thebibliography}
\end{document}